\documentstyle[12pt]{article}

\newcounter{saveeqn}
\newcounter{App} 
\newcommand{\app}{%
\stepcounter{App}%
\setcounter{saveeqn}{\value{equation}}%
\setcounter{equation}{0}%
\renewcommand{\theequation}{\Alph{App}\arabic{equation}} }
\newcommand{\appende}{%
\setcounter{equation}{\value{saveeqn}}%
\renewcommand{\theequation}{\arabic{equation}}  }

\def\be{\begin{equation}}
\def\ee{\end{equation}}
\def\bq{\begin{eqnarray}}
\def\eq{\end{eqnarray}}
\def\g{\gamma}
\def\ra{\rightarrow}
\def\ve{\varepsilon}
\def\vp{\varphi}
\def\n{\nonumber}
\def\bu{\bar{u}}
\def\gmr{g_{\mu\rho}}
\def\gnr{g_{\nu\rho}}
\def\gmn{g_{\mu\nu}}
\def\qn{q_\nu}
\def\qm{q_\mu}
\def\qr{q_\rho}
\def\xn{x_\nu}
\def\xm{x_\mu}
\def\xr{x_\rho}
\def\gnl{g_{\nu\lambda}}
\def\gml{g_{\mu\lambda}}
\def\grl{g_{\rho\lambda}}

\begin{document}

\begin{flushright}
CEBAF-TH-94-09
\end{flushright}
\vspace{2cm}
\begin{center}
{\bf $\pi$-$A_1$ electromagnetic form factors\\ 
and light-cone QCD sum rules.}
\end{center}
\begin{center}{V.M. Belyaev$^*$}
\end{center}
 \begin{center} 
 Continuous Electron Beam Accelerator Facility  
12000 Jefferson Ave, Newport News, Virginia 23606, USA
\end{center}
\vspace{2cm}
\begin{abstract}
Electromagnetic form factors of the transition $\pi+\gamma_{virt.}\ra A_1$ 
are calculated by QCD sum rules technique with the description of the pion
in terms of the set of wave functions of increasing twist. Obtained results
are compared with standard QCD sum rule calculations.
\end{abstract}
\vspace{5cm}
\flushbottom{$^*\overline{On\;  leave\;  of\;  absence }
\;from\;  ITEP,\;  117259\;  Moscow,\; Russia.$}

\newpage

\section*{}

Recently  it was suggested to study pion form factor at not very large 
momentum transfers
by light-cone QCD sum rules  \cite{braun} which combine the  description of 
pion in terms of the set of 
wave functions of increasing twist with the technique of QCD sum rules 
\cite{SVZ}. This approach
gives a possibility to calculate a contribution of so-called Feynman 
mechanism  to the
pion form factor, in which large momentum transfer selects the configuration, in which one 
parton carries almost
all the momentum of the hadron. In that paper  \cite{braun} it was shown 
that at least up to the momentum
transfers of order $10$ $GeV^2$ this mechanism remains important and it is 
possible to describe pion form
factor in this region by Feynman mechanism only. The second mechanism of 
large momentum transfer
is the hard rescattering mechanism, in which large momentum transfer selects 
configurations with a small transfers
size in which the momentum fraction carried by interacting quark remains an 
average one.
The hard rescattering mechanism involves a hard gluon exchange, and can be 
written in the factorized
form \cite{cz}, \cite{er}, \cite{lb}.

There exists a number of arguments in favour of  that at large but finite 
momentum transfer ($Q^2\sim 1 - 10GeV^2$)
 Feynman mechanism is dominant.  Using QCD sum rule approach \cite{SVZ} it 
was found \cite{i1},\cite{r} that
pion form factor at $Q^2\sim 1-3 GeV^2$ is saturated by Feynman type 
contribution. However, this method
can not be  used for higher momentum transfer due to increasing  contribution 
of higher operator
product expansion terms at large $Q^2$. Nevertheless in \cite{br},\cite{mr} 
using a concept  of nonlocal
condensates it was obtained indications that Feynman type contribution is 
dominate at least up to
$Q^2\sim 10 GeV^2$.  At the same time an attempt to describe the data of 
pion form factor at $Q^2\geq 3GeV^2$
by the contribution of hard scattering only leads to conclusion that the low 
energy pion wave function
is very far from its asymptotic form \cite{cz2} and there was proposed a 
model for pion wave function
which has a peculiar "humped" profile which corresponds that the most 
probable pion configuration is
the case when one of the quark carry almost all pion momentum. But in 
\cite{isgur} was pointed that
the wave function of such type corresponds to respectively soft gluon 
exchange even at $Q^2\sim 10GeV^2$.

Here we discuss Feynman type mechanism for $\pi\ra A_1$ 
electromagnetic pion form factors. The first  calculation
of this amplitude was made in \cite{i2} by using operator product expansion 
for three-points correlator in vacuum.

\section{}
 Let us consider the correlator:
\bq
T_{\mu\nu}(p,q)=\int e^{ipx}d^4x<0|T\{ j_\mu^5(x),j_\nu^{em.}(0)\} |\pi (k)>,
\label{1}
\eq
where $j_\mu^5=\bar{d}\g_\mu\g_5u$ and $j_\nu^{em.}=\frac23 \bar{u}\g_\nu 
u-\frac13 \bar d\g_\nu d$ is the
electromagnetic current, $k$ is momentum of pion.  This correlator was used 
in \cite{braun} to study
pion form factor.

Leading twist operator gives the following contribution to $T_{\mu\nu}$:
\bq
T_{\mu\nu}(p,q)=f_\pi\int_0^1du\frac{\vp_\pi(u)}{(1-u)p^2+uq^2}\left(\frac12
(p^2-q^2)g_{\mu\nu}
-2(1-u)p_\mu p_\nu\right.
\n
\\
\left.+(1-2u)(p_\mu q_\nu+q_\mu p_\nu)+2uq_\mu q_\nu\right)
\label{2}
\eq
where $\vp_\pi(u)$ is twist-2 pion wave function. We use the following 
definition for twist$-2$
and $-4$ two-particle wave functions of pion:
\bq
<0|\bar d(0)\g_\mu\g_5u(x)|\pi(k)>=if_\pi k_\mu\int_0^1e^{-iukx}(\vp(u)+
x^2g_1(u)+O(x^4))du
\n
\\
+f_\pi\left(x_\mu-\frac{k_\mu}{kx}x^2\right)\int_0^1e^{-iukx}g_2(u)du+...
\label{3}
\eq
Here $g_1$ and $g_2$ are twist-$4$ pion wave functions.

In this paper we use  different models for pion wave function.
The first on is asymptotical wave function.
In \cite{cz} it was shown that at asymptotically large $Q^2$ the pion wave 
function of leading twist
has the following form:
\bq
\vp_\pi^{(as.)}(u)=6u(1-u)
\label{3a}
\eq
The attempt to describe pion form factor at $Q^2\geq 3 GeV^2$ by hard 
rescattering mechanism only
leads to a conclusion that the form of the wave function is much different 
from asymptotical one
and it was suggested to use the following model for pion twist-2 wave 
function (see \cite{cz2}):
\bq
\vp_\pi^{(CZ)}(u, \mu\sim 500 MeV)=30u(1-u)(2u-1)^2
\label{3b}
\eq
And the third model pion wave function is
\bq
\vp_\pi^{(BF)} (u)=6u(1-u)\left(1+A_2\frac32[5(2u-1)^2-1]\right.
\n
\\
\left.+A_4\frac{15}8[21(2u-1)^4-14(2u-1)^2+1]\right)
\label{3c}
\eq
which was proposed by Braun and Filyanov in \cite{bf} at low normalization 
point $\mu\simeq  0.5 GeV$, with
coefficients $A_2=\frac23$ and $A_4=0.43$ which is in agreement with QCD sum 
rules for
first moments of pion wave function and provides that at a middle point 
$u=0.5$
\bq
\vp_\pi(0.5)\simeq 1.2.
\label{3d}
\eq
This value is with experimental values of various hadronic coupling 
constants calculated
be light-cone sum rules approach (see \cite{bf}).

It is easy to check that (\ref{2}) satisfies to Ward Identities:
\bq
p_\mu F_{\mu\nu}=-f_\pi (p-q)_\nu
\n
\\
q_\nu F_{\mu\nu}=-f_\pi (p-q)_\mu.
\label{4}
\eq

Twist-$4$ quark wave functions $g_1(u)$ and $g_2(u)$ give the following 
contribution into the
correlator (\ref{1}):
\bq
f_\pi\int_0^1du\left(\frac{1}{((1-u)p^2+uq^2)^2}\left[4g_1(u)(\frac12 
(p^2-q^2)g_{\mu\nu}\right.\right.
\n
\\-2(1-u)p_\mu p_\nu+(p_\mu q_\nu+q_\mu p_\nu)(1-2u)+2uq_\mu q_\nu)
\n
\\
-4g_2(u)((1-u)^2p_\mu p_\nu+u(1-u)(p_\mu q_\nu+q_\mu p_\nu)+u^2q_\mu q_\nu)
\n
\\
\left.
-4{\cal G}_2(u)(2(1-u)p_\mu p_\nu-(1-2u)(p_\mu q_\nu+q_\mu p_\nu)-2uq_\mu q_\nu)
\right]
\n
\\
\left.
+\frac{2g_2(u)}{(1-u)p^2+uq^2}g_{\mu\nu}\right),
\label{5}
\eq
where ${\cal G}_2(u)=-\int_0^ug_2(v)dv$.

There are also quark-gluon twist-$4$ operators:
\bq
<0|\bar d(0)\g_\mu\g_5g_sG_{\alpha\beta}(vx)u(x)|\pi(k)>
\n
\\
=k_\mu(k_\alpha x_\beta-x_\alpha k_\beta)\frac1\pi
f_\pi\int e^{-ikx(\alpha_1+v\alpha_3)} D\alpha\Phi_\parallel(\alpha_i)
\n
\\
+\left(k_\beta(g_{\mu\alpha}-\frac{k_\mu  x_\alpha}{kx})-k_\alpha
(g_\mu\beta-\frac{k_\mu x_\beta}{kx})\right)
f_\pi\int e^{-ikx(\alpha_1+v\alpha_3)} D\alpha\Phi_\perp(\alpha_i),
\label{6}
\eq
\bq
<0|\bar d(0)\g_\mu ig_s\tilde{G}_{\alpha\beta}(vx)u(x)|\pi(k)>
\n
\\
=k_\mu(k_\alpha x_\beta-x_\alpha k_\beta)\frac1\pi
f_\pi\int e^{-ikx(\alpha_1+v\alpha_3)} D\alpha\Psi_\parallel(\alpha_i)
\n
\\
+\left(k_\beta(g_{\mu\alpha}-\frac{k_\mu  x_\alpha}{kx})-k_\alpha(g_\mu\beta-
\frac{k_\mu x_\beta}{kx})\right)
f_\pi\int e^{-ikx(\alpha_1+v\alpha_3)} D\alpha_i\Psi_\perp(\alpha_i),
\label{7}
\eq
where $\tilde{G}_{\alpha\beta}=\frac12 \ve_{\alpha\beta\mu\nu}G^{\mu\nu}$, 
$D\alpha_i=d\alpha_1d\alpha_2
d\alpha_3\delta(1-\alpha_1-\alpha_2-\alpha_3)$.

The contribution of these operators (\ref{6},\ref{7}) is:
\bq
f_\pi\int_0^1du\frac{1}{((1-u)p^2+uq^2)^2}\left[(2(1-u)p_\mu p_\nu-(1-2u)
(p_\mu q_\nu+q_\mu p_\nu)\right.
\n
\\
\left.-2uq_\mu q_\nu)A(u)-\frac12 g_{\mu\nu}(p^2-q^2)B(u)+2(p_\mu q_\nu-
q_\mu p_\nu)C(u)\right],
\label{8}
\eq
where
\bq
A(u)=\int_0^ud\alpha_1\int_{u-\alpha_1}^{1-\alpha_1}\frac{d\alpha_3}{\alpha_3}
\left[
\Psi_\parallel (\alpha_i)+2\Psi_\perp (\alpha_i)\right.
\n
\\
\left. +\left(1-2\frac{u-\alpha_1}{\alpha_3}\right)(\Phi_\parallel (\alpha_i)
+2\Phi_\perp (\alpha_i))\right],
\label{9}
\eq
\bq
B(u)=\int_0^ud\alpha_1\int_{u-\alpha_1}^{1-\alpha_1}\frac{d\alpha_3}
{\alpha_3}\left[\Psi_\parallel (\alpha_i)
+\left(1-2\frac{u-\alpha_1}{\alpha_3}\right)\Phi_\parallel (\alpha_i)
\right],
\label{10}
\eq
\bq
C(u)=\int_0^ud\alpha_1\int_{u-\alpha_1}^{1-\alpha_1}\frac{d\alpha_3}
{\alpha_3}\left[
\Phi_\perp (\alpha_i)+\left(1-2\frac{u-\alpha_1}{\alpha_3}\right)
\Psi_\perp(\alpha_i)\right].
\label{11}
\eq
There are also 4-particle twist-4 operators which were considered in
ref.\cite{gorsky}. Contributions of such operators are not considered
here.

A systematic study of the higher twist wave functions 
( excluding four-particle operators)       has been done in the
paper \cite{bf}.
The set of wave functions suggested in the paper includes contributions of 
operators with lowest
conformal spin and also the  corrections corresponding to the operators with
 next-to-leading
conformal spin, which numerical values were estimated by the QCD sum rule 
method. This set
is (hereafter $\bar{u}=1-u$):
\bq
\Phi_\parallel(\alpha_i)&=&120 \ve\delta^2(\alpha_1 -\alpha_2 )\alpha_1 
\alpha_2 \alpha_3,
\n
\\
\Psi_\parallel(\alpha_i)&=&-120\delta^2\alpha_1\alpha_2\alpha_3\left(\frac13
+\ve(1-3\alpha_3)\right),
\n
\\
\Phi_\perp(\alpha_i)&=&30\delta^2(\alpha_1-\alpha_2)\alpha_3^2\left(\frac13
+2\ve(1-2\alpha_3)\right),
\n
\\
\Psi_\perp(\alpha_i)&=&30\delta^2(1-\alpha_3)\alpha_3^2\left(\frac13+2\ve
(1-2\alpha_3)\right),
\label{11a}
\\
g_1(u)&=&\frac52 \delta^2\bu^2u^2+\frac1\ve\delta^2\left(\bu u(2+13\bu u)+
10u^3(2-3u+\frac65u^2)
\ln (u)\right.
\n
\\
&&\left.+10\bu^3(2-3\bu+\frac65\bu^2)
\ln (\bu)\right),
\n
\\
g_2(u)&=&\frac{10}3\delta^2\bu u(u-\bu),
\n
\\
{\cal G}_2(u)&=&\frac53\delta^2\bu^2 u^2,
\eq
where
\bq
\delta^2&\simeq& 0.2GeV^2,\;\;\;\ve\simeq 0.5.
\label{11bb}
\eq
The value of $\delta^2$ was determined in \cite{n}.

Using this set of the wave function we obtain the following expressions for 
$A(u)$, $B(u)$ and
$C(u)$:
\bq
A(u)&=&\delta^2\left( 10u^2(1-u)^2+\ve\left[ -4u-22u^2+52u^3-26u^4-4\ln 
(1-u)
\right.\right.
\n
\\
&&\left.\left.+4u^3(10-15u+12u^2)\ln\left(\frac{1-u}u\right)\right]\right),
\label{11b}
\\
B(u)&=&\delta^2\left(-10u^2(1-u)^2+\ve\left[-4u-22u^2+52u^3-26u^4-4\ln 
(1-u)\right.\right.
\n
\\
&&\left.\left.+4u^3(10-15u+12u^2)\ln\left(\frac{1-u}u\right)\right]\right),
\label{11c}
\\
C(u)&=&20\delta^2\ve u^2(1-u)^2(2u-1).
\label{11d}
\eq

\section{}
Let us consider hadron contribution into correlator $T_{\mu\nu}$.
$A_1$ -meson gives the following contribution into this correlator:
\bq
\frac{-i}{m^2_A-p^2}<0|j_\mu^5|A_1(p)><A_1(p)|j_\nu^{em.}|\pi (k)>
\n
\\
=\frac{f_A}{m^2_A-p^2}\left[g_{\mu\nu}m_A^2G_1(Q^2)-p_\mu p_\nu\frac{Q^2}
{m^2_A}\left(G_1(Q^2)-
\left(1-\frac{Q^2}{m_A^2}\right)G_2(Q^2)\right)\right.
\n
\\
-p_\mu q_\nu\frac12\left(1-\frac{Q^2}{m_A^2}\right)\left(G_1(Q^2)-\left(1-
\frac{Q^2}{m_A^2}\right)
G_2(Q^2)\right)
\n
\\
-2q_\mu p_\nu \left( G_1(Q^2)+\frac{Q^2}{m_A^2}G_2(Q^2)\right)
\n
\\
\left. +q_\mu q_\nu\left( G_1(Q^2)-\left(1-\frac{Q^2}{m_A^2}\right)G_2(Q^2)
\right)\right],
\label{12}
\eq
where
\bq
<0|j_\mu^5|A_1(p)>=i\epsilon_\mu m_A f_A,
\label{13}
\eq
\bq
<A_1|j_\nu^{em.}|\pi(k)>=-\frac{\epsilon_\lambda}{m_A}\left[ (2p-q,q)
g_{\nu\lambda}-(2p-q)_\nu
q_\lambda)G_1(Q^2)\right.
\n
\\
\left.-\frac{1}{m_A^2}((2p-q,q)q_\nu-q^2(2p-q)_\nu)q_\lambda G_2(Q^2)\right],
\label{14}
\eq
$m_A=1.26GeV$ is $A_1$-meson mass, $\epsilon_\lambda$ is a polarization of 
the meson,
$f_A=0.2$ which was determined in \cite{fa} from a fit to the absolute rate 
for
$\tau\ra \nu_\tau\pi\pi\pi$ decay.
Notice that in paper
\cite{i2} it was used another set for mass and $A_1$-coupling
constant -  $\frac{g^2_{A_1}}{4\pi}\simeq 6.0$ and $m_A=1.15  GeV$.
This parameters were determined in
\cite{ga} by QCD sum rule, $f_A=\sqrt{2}\frac{m_A}{g_{A_1}}$.

The heavier spin-$1$ states contributions have the same form as
 (\ref{12}).

In the limit of massless quarks due to the conservation of axial current 
$j_\mu^5$, 
the only massless state of spin-$0$ (pion) gives nonvanishing contribution 
into
the correlator (\ref{1}). This contribution is
\bq
\frac{i}{p^2}<0|j_\mu^5|\pi(p)><\pi(p)|j_\nu^{em.}|\pi(k)>
=-\frac{f_\pi p_\mu}{p^2}F_\pi(Q^2)(2p-q)_\nu.
\label{15}
\eq
Pion  form factor $F_\pi(Q^)$  was studied in \cite{braun}.

\section{}
To study $G_1$ form factor we consider correlation function $f_1(p^2,q^2)$:
\bq
T_{\mu\nu}(p,q)=g_{\mu\nu}f_1(p^2,q^2)+...(other \;\;tensors).
\label{16}
\eq
Then according to (\ref{10}) and (\ref{15}) it is possible to 
write the following dispersional relation for the
structure $g_{\mu\nu}$:
\bq
f_1(p^2,q^2)=\frac{f_Am_A^2G_1(Q^2)}{m^2_A-p^2}+\frac{1}{\pi}\int_{s_0}^{
\infty}
\frac{\rho(s,Q^2)ds}{p^2-s}+(subtraction\;terms),
\label{17}
\eq
where $\rho(s,Q^2)$ is  the spectral density of higher state (spin-$1$) 
contribution
into the correlator (\ref{1}), $s_0>m_A^2$ and pion contribution is absent.

To suppress higher states contribution and to kill subtraction terms in 
(\ref{17}) we
use so called Borel operator ${\cal B}$ which was suggested in \cite{SVZ}:
\bq
{\cal B}_{P^2}f(P^2)=\frac{(P^2)^{(n+1)}}{n!}\left(-\frac{d}{dP^2}\right)^{n}
f(P^2)=f^B(t),
\n
\\
\frac{P^2}{t}=n,\;\;\;\;P^2=-p^2,\;\;\;\; n\ra\infty.
\label{18}
\eq

Applying this operator to (\ref{17}) we obtain
\bq
f_1^B(t,Q^2)=f_Am_A^2G_1(Q^2)e^{-\frac{m_A^2}{t}}+\frac{1}{\pi}\int_{s_0}^{
\infty}
e^{-\frac{s}{t}}\rho(s,Q^2)ds.
\label{19}
\eq
Notice, that higher states contribution is suppressed by a factor $e^{-s/t}
\leq e^{-s_0/t}$.

From other side, the leading twist-2  wave function  gives the following 
contribution to
$f_1^B$:
\bq
f_1^{B(twist-2)}=\frac{f_\pi}2\int_0^1 \frac{Q^2\vp_\pi(u)}{(1-u)^2}e^{-
\frac{uQ^2}{(1-u)t}}du
\n
\\
=\frac{f_\pi}2\int_0^\infty\vp_\pi\left(\frac{s}{s+Q^2}\right)e^{-\frac{s}t}
ds.
\label{20}
\eq

In (\ref{20}) we represent the formula in the form of dispersional integral 
after Borel transformation and
$\frac{f_\pi}2\vp_\pi\left(\frac{s}{s+Q^2}\right)$ is equal  to imaginary 
part of the correlator
at $p^2=s$. 

According to (\ref{5}) and (\ref{8})  twist-$4$ wave functions contributions 
are
\bq
f_1^{B(twist-4)}=\frac{f_\pi}2\int_0^1\left(\frac{Q^2(4g_1(u)-B(u))}
{t(1-u)^3}-\frac{4g_1(u)-B(u)}{(1-u)^2}\right.
\n
\\
\left.- \frac{4g_2(u)}{(1-u)}\right)e^{-\frac{uQ^2}{(1-u)t}}du
=\frac{f_\pi}2\int_0^\infty ds\left(\left[4g_1\left(\frac{s}{s+Q^2}\right)
\right.\right.
\n
\\
\left.\left.
-
B\left(\frac{s}{s+Q^2}\right)\right]\left[\frac{s+Q^2}{Q^2t}-\frac1{Q^2}
\right]
-
\frac{4 g_2\left(\frac{s}{s+Q^2}\right)}{s+Q^2}\right)e^{-\frac{s}t}ds.
\label{21}
\eq

Due to asymptotic  freedom in the limit $s\ra\infty$ the imaginary part of 
the correlator
tends to his perturbative value. So it is possible, like in usual sum rules, 
to estimate
higher states contribution   suggested that the imaginary part of the 
 correlator
(\ref{1}) is equal to its calculable part (by operator product expansion) 
starting at
$s=s_0$. Or by other words,  here we use the standard concept of duality, 
which tell us
that $A_1-$meson occupies the "region of duality" in the invariant mass $s$ 
up to
a certain threshold $s_0\simeq 3 GeV^2$. Thus, to take into account higher 
states
contribution we use the following limits in of  integration 
(\ref{20},\ref{21}):
\bq
0 \leq u\leq \frac{s_0}{s_0+Q^2}\;\; or \;\;0\leq s\leq s_0.
\label{25}
\eq

Using eqs.(\ref{19}-\ref{25}) we obtain the following sum rule:
\bq
G_1(Q^2)=\frac{f_1^{B(twist-2)}+f_1^{(B(twist-4)}}{f_A m_A^2}
e^{\frac{m_A^2}t},
\label{31}
\eq
where we use a standard model for higher state contribution using the limits 
of integration
(\ref{25}) in expressions for $f_1^{B(2)}$ and $f_1^{B(4)}$.

Three models of the leading twist wave function $\vp_\pi(u)$ were considered.
In Fig.1 it was shown $Q^2-$dependence of $G_1$ form factor which was
calculated using eq.(\ref{31}) at $t=1.5GeV^2$ and $s_0=3GeV^2$.

Notice, that the dependence of $G_1$ on the pion wave function is very weak 
at $Q^2\sim 2.5GeV^2$.
The reason of this is that at $Q^2\sim s_0=3GeV^2$ we integrate a pion wave 
function in the region
$0<u<0.5$. In the limit $t\ra\infty$ this integral is equal to $0.5$ and 
does not depend on the form
of wave function.  It is clear that using for respectively large $t$ the 
dependence of sum rule  on the
pion wave function will be weak and at $Q^2\simeq 2.5 GeV^2$ this dependence 
is compensated by
a small changing of integration region over $u$. 

In this picture we show predictions which were obtained  by Ioffe and Smilga 
in
\cite{i2} from QCD sum rules for three-point correlator.
Notice that there is a big disagreement between predictions of QCD sum 
rules
for three-point correlator and the case  of asymptotic wave function for 
$\pi$-meson.
The best agreement with predictions  of Ioffe and Smilga at 
$Q^2\leq 2GeV^2$
 we have in the case of
BF-wave function.  
At $Q^2\sim 3GeV^2$ the predictions of \cite{i2} two times smaller
than the result obtained in this paper. Probably, this disagreement can be 
explained
by large contribution of higher dimensions operators which becomes large at
this momentum transfer (see details in \cite{i2}). 
At $Q^2\leq 2GeV^2$ disagreement between
the case of BF-wave function and three-point sum rule is about 20\% which
is usual accuracy of QCD sum rules.
In this picture we show the experimental value for $G_1(0)$ which is 
determined
from partial width of decay $A_1\ra\pi\g$. 
According to \cite{z} $\Gamma(A_1\ra\pi\g)=640\pm 246 keV$.
 Stability of the sum rule (\ref{31}) is illustrated in Fig.2.
Notice that at $Q^2\geq 15 GeV^2$ continuum contribution becomes larger
than 30\%. Twist$-4$ operator contribution is smaller than 10\%.

\section{}
To find $G_2$ form factor, let us consider $q_\mu p_\nu$ and
$q_\mu q_\nu$  tensor structures of contribution of hadrons into the 
correlator
(\ref{1}).  The only spin$-1$ states give nonzero contribution into these 
structures.
In the case $q_\mu p_\nu$ this contribution has the following form:
\bq
f_\alpha(p^2,Q^2)=
-\sum_i\frac{2 f_i}{m_i^2-p^2}\left(G_1^{(i)}(Q^2)+\frac{Q^2}{m^2_i}G_2^{(i)}
(Q^2)\right),
\label{32}
\eq
and  in the case $q_\mu q_\nu$ the contribution is:
\bq
f_\beta(p^2,Q^2)=\sum_i\frac{f_i}{m_i^2-p^2}\left(\left(G_1^{(i)}(Q^2)+
\frac{Q^2}{m_i^2}G_2^{(i)}(Q^2)\right)-
G_2^{(i)}(Q^2)\right),
\label{33}
\eq
where $\sum_i$ is a sum over spin-1 resonances ($A_1$ and higher states with
$J^{PC}=1^{+-}$, $m_i$ and $f_i$ are their masses and residues into axial 
current,
$G_1^{(i)}$ and $G_2^{(i)}$ are their form factors. 
From other side, $f_\alpha$ and $f_\beta$ were calculated in
Section II. Thus using eqs.(\ref{32},\ref{33}) we have the following sum 
rule for
$G_2$ form factor:
\bq
-\frac12f_\alpha(p^2,Q^2)-f_\beta(p^2,Q^2)=
\sum_i\frac{f_i}{m_i^2-p^2}G_1^{(i)}(Q^2).
\label{34}
\eq

The left side of this expression was found in Section II at $p^2\sim 1-3 
GeV^2$ 
and $Q^2>1GeV^2$
Using Borel operator (see (\ref{18})) and using the model of continuum which 
was described in
previous Section to take into account higher state
contributions we obtain the following sum rule:
\bq
\frac{f_\pi}{f_A}e^{\frac{m_A^2}t}\int_0^{u_0}e^{-\frac{Q^2u}{t(1-u)}}
du\left\{
\frac{(\frac12+u)\vp_\pi(u)}{1-u}\right.
\n
\\
\left.+\frac1{(1-u)^2t}\left( (1+2u)(-2g_1(u)-2{\cal G}_2(u)\right.\right.
\n
\\
\left.\left.+\frac12A(u))+2u(1+u)g_2(u)
+C(u)\right)\right\}
\label{35}
\eq
\bq
&=&\frac{f_\pi}{f_A}e^{\frac{m_A^2}t}\int_0^{s_0}\left\{
\frac{(3s+Q^2)\vp_\pi\left(\frac{s}{s+Q^2}\right)}{2(s+Q^2)^2}\right.
\n
\\
&&+\frac1{Q^2 t}\left[ \frac{3s+Q^2}{s+Q^2}\left(
-2g_1\left(\frac{s}{s+Q^2}\right)-2{\cal G}_2\left(\frac{s}{s+Q^2}\right)
+\frac12A\left(\frac{s}{s+Q^2}\right)\right)\right.
\n
\\
&&\left.\left.+2\frac{s(2s+Q^2)}{(s+Q^2)^2}g_2\left(\frac{s}{s+Q^2}\right)
+C\left(\frac{s}{s+Q^2}\right)\right]\right\}.
=G_2(Q^2)
\label{36}
\eq
$Q^2$ dependence of $G_2$ form factor for different wave functions 
are shown in Fig.3.  These results were obtained from sum rule
(\ref{36}) at $t=1.5GeV^2$ and $s_0=3GeV^2$.
It is interesting compare the results obtained with the first calculation 
of this formfactor which was made by Ioffe and Smilga \cite{i2}.
They used QCD sum rules for three point correlator
and their results are depicted in Fig.3 too.
In the case of this form factor there is a good agreement between the case 
of asymptotic wave function
and predictions which was obtained in \cite{i2}. We have a big disagreement 
between 
our results for CZ and BF wave functions at $Q^2\sim 1GeV^2$.
This disagreement can be explained by a large contribution of higher twist 
operators.
 At $Q^2>2GeV^2$ the contribution of twist-4 operators
becomes smaller 20\%. At lower $Q^2\sim 1 GeV^2$ this contribution is more 
than 30\%.
Stablity of the sum rules is illustrated in Fig.4.

\section{}
In this paper we have used light-cone sum rules to calculate electromagnetic
$\pi-A_1$ form factors in the region of intermediate momentum transfers:
$1GeV^2<Q^2<15GeV^2$.  
It was studied the dependence of the results obtained on
the form  of the pion wave function of leading twist.
It was found that at $Q^2>3GeV^2$ the behaviour of the form factors in the 
case
of asymptotical wave function are very different from the cases of CZ and 
BF wave
functions.  
It was shown that the results obtained in the cases of CZ and BF
wave functions are not very different from each other at $Q^2\sim 10 GeV^2$.

We compare the results obtained in this paper
with the first calculations of these form factors
which was made in framework of QCD sum rules for three point correlator by
Ioffe and Smilga \cite{i2}. These two predictions are in agreement within 
20-30\%
accuracy at $Q^2\sim 1-2GeV^2$. 
This accuracy is usual
accuracy of QCD sum rules. It is shown that there are
 disagreements between these two approaches
at $Q^2>2GeV^2$ and $Q^2<1GeV^2$ which
can be explained by large contribution of power corrections at high $Q^2$
in the case of three-points sum rule and by contribution of higher twist 
operators 
at respectively small $Q^2$ in the case of light-cone sum rules.

We obtain that vector dominance does not work in the process.
If we try to fit the form factors  by $\rho$ and $\rho '$-mesons
then we find that $\rho '$-meson contribution is dominant at $Q^2>1GeV^2$,
and we can not use vector dominance to estimate $\rho -A_1-\pi$-coupling 
constant.

It is shown that at $Q^2>1-2GeV^2$ form factor $G_1$ increases with
growth of $Q^2$. 
This behaviour can be explained in framework of naive quark
model, where $A_1$-meson is consist of two quarks in $P$-wave.

Notice that it will be very interesting to measure $G_1$ form factor at 
$Q^2\sim 2.5GeV^2$
because here we have a very weak dependence on the form of pion wave function.
The slope of the form factor  gives additional information on the function.
\\ \\ 
{\it Acknowledgements.} The author is grateful to V.M. Braun, N. Isgur,  and A.V.  
Radyushkin
for useful discussions. 

\section*{Appendix A}
\app

In \cite{bf} it was shown that there are relations between wave functions of 
the same twist.
Here one more relation for twist-4 wave function will be obtain.
Notice, that a set of wave functions suggested in \cite{bf} is in agreement 
with this relation.
It is a generalization of one of the relations obtained in \cite{bf}.

Let us consider the following matrix element:
\bq
\frac1{f_\pi}<0|\bar{d}(0)\g_\mu\g_\rho\g_\nu\g_5 u(x)|\pi>=
(\gmr\qn+\gnr\qm-\gmn\qr)
\nonumber
\\
\times\int_0^1due^{-iuqx}\left(i\vp_\pi(u)+
ix^2g_1(u)-\frac{x^2}{(qx)}g_2(u)\right)
\nonumber
\\
+(\gmr\xn+\gnr\xm-\gmn\xr)\int_0^1due^{-iuqx} g_2(u),
\label{a1}
\eq
and find first derivative of (\ref{a1}) over $\xn$ 
\bq
i\gmr(qx)\int_0^1due^{-iuqx}(2g_1(u)+2{\cal G}_2(u)-ug_2(u))
\nonumber
\\
+i(\qm\xr-\qr\xm)\int_0^1due^{-iuqx}(2g_1(u)+2{\cal G}_2(u)+ug_2(u)).
\label{a2}
\eq
From other side, using equation of motion for massless quarks we
have:
\bq
\frac1{f_\pi}\frac{\partial}{\partial x_\nu}
<0|\bar{d}(0)\g_\mu\g_\rho\g_\nu\g_5 u(x)|\pi>
\nonumber
\\
=i\frac1{f_\pi}\int_0^1vdv<0|\bar{d}(0)\g_\mu\g_\rho\g_\nu\g_5 
x_\alpha gG_{\alpha\nu}(vx)u(x)|\pi>
\label{a3}
\eq
\bq
=i\frac1{f_\pi}\int_0^1vdvx_\alpha\left( (\gmr\gnl+\gnr\gml-\gmn\grl)
\right.
\n
\\
\left. <0|\bar{d}(0)gG_{\alpha\nu}(vx)\g_\lambda\g_5u(x)|\pi>+\ve_{
\mu\nu\rho\lambda}
<0|\bar{d}(0)igG_{\alpha\nu}\g_\lambda u(x)|\pi>\right)
\label{a5}
\eq
\bq
=i\gmr (qx)\int_0^1vdv{\cal D}\alpha_ie^{-iqx(\alpha_1+v\alpha_3)}\left(
\Phi_\parallel(\alpha_i)-
2\Phi_\perp (\alpha_i)\right)
\n
\\
+i(\qm\xr-\xm\qr)\int_0^1vdv{\cal D}\alpha_ie^{-iqx(\alpha_1+v\alpha_3)}
\left( \Phi_\parallel(\alpha_i)
+2\Psi_\perp(\alpha_i)\right).
\label{a7}
\eq
Using new variable $u$:
\bq
\alpha_1+v\alpha_3=u\;\;\;v=\frac{u-\alpha_1}{\alpha_3}
\label{a8}
\eq
eq.(\ref{a7}) becomes
\bq
i\gmr (qx)\int_0^1e^{-iqxu}du {\cal A}(u)+i(\qm\xr-\xm\qr)\int_0^1
e^{-iqxu}du {\cal B(u)},
\label{a9}
\eq
where
\bq
{\cal A}(u)=\int_0^ud\alpha_1\int_{u-\alpha_1}^{1-\alpha_1}
\frac{d\alpha_3}{\alpha_3^2}(u-\alpha_1)\left(
\Phi_\parallel(\alpha_i)-
2\Phi_\perp (\alpha_i)\right),
\label{a10}
\\
{\cal B}(u)=\int_0^ud\alpha_1\int_{u-\alpha_1}^{1-\alpha_1}
\frac{d\alpha_3}{\alpha_3^2}(u-\alpha_1)\left(
\Phi_\parallel(\alpha_i)
+2\Psi_\perp(\alpha_i)\right).
\label{a11}
\eq
Comparing eqs.(\ref{a2}) and (\ref{a9}) we obtain:
\bq
2g_1(u)&-&2{\cal G}_2(u)-ug_2(u)
\n
\\
&=&\int_0^ud\alpha_1\int_{u-\alpha_1}^{1-\alpha_1}
\frac{d\alpha_3}{\alpha_3^2}(u-\alpha_1)\left(
\Phi_\parallel(\alpha_i)-
2\Phi_\perp (\alpha_i)\right),
\label{a12}
\eq
\bq
2g_1(u)&+&2{\cal G}_2(u)+ug_2(u)
\n
\\
&=&\int_0^ud\alpha_1\int_{u-\alpha_1}^{1-\alpha_1}
\frac{d\alpha_3}{\alpha_3^2}(u-\alpha_1)\left(
\Phi_\parallel(\alpha_i)
+2\Psi_\perp(\alpha_i)\right).
\label{a13}
\eq
The first relation (\ref{a12}) was obtained in \cite{bf}. The second 
equation (\ref{a13}) is a new
one.  Using (\ref{a12}) and (\ref{a13}) we obtain
\bq
2{\cal G}_2(u)+ug_2(u)=
\int_0^ud\alpha_1\int_{u-\alpha_1}^{1-\alpha_1}
\frac{d\alpha_3}{\alpha_3^2}(u-\alpha_1)\left(
\Phi_\perp(\alpha_i)
+\Psi_\perp(\alpha_i)\right).
\label{a14}
\eq
It is  not difficult to check that the 
set of wave functions suggested in \cite{bf} satisfies to this new relation 
(\ref{a14}).
Notice that the set of wave functions was fixed
by using two integral relations which are valid for any twist-4 wave 
functions, and  additional one
 which was obtained for asymptotic and preasymptotic twist-4 wave functions. 
Thus, this new integral relation is a generalization of the constraints 
obtained in \cite{bf}
for the case of asymptotic and preasymptotic wave functions.

Let we write the last relation for twist-4 wave function which was obtained 
in \cite{bf}:
\bq
g_2(u)=\int_0^ud\alpha_1\int_{u-\alpha_1}^{1-\alpha_1}\frac{d\alpha_3}
{\alpha_3}
\left(2\Phi_\perp(\alpha_i)-\Phi_\parallel(\alpha_i)\right)
\label{a15}
\eq

Equation  (\ref{a14}) can be useful for construction of a set of wave 
functions which are far from
 asymptotic ones.
\appende

\end{document}